# A Tribute to Bill Kruskal

**Norman M. Bradburn**

Bill Kruskal was a dedicated citizen of the University of Chicago, of the broader statistical community and of the country.

It is difficult to convey fully Bill's commitment to and contributions to the community life of the University. The University of Chicago, unusual among American universities, believes that the maintenance and enhancement of a scholarly community is part of its mission. Maintaining a true scholarly community is difficult and rests heavily on those who are willing to commit some of their time and energy to the betterment of the community. Bill was an extraordinarily good citizen, He chaired a number of faculty committees that dealt with some of the most delicate issues of university life, notably the Committee on Appointment Inequities; he was a member of the Council of the University Senate and of the Committee of the Council; he served as chairman of the Statistics Department and in a bold move, for he was not a member of its faculty, served as dean of the Social Sciences Division for two terms, as well as serving as dean pro tempore during the Harris School's inaugural year.

This concern for the University extended to the practices of the library. When he could not find a particular issue of a Census Bureau serial publication, but instead found a note taped in the bound volume saying that the library had discarded number 8 at the request of the Government Printing Office because the report had too many mistakes, he wrote in some indignation, "How could our library discard a document that ipso facto held such sociological, ethical, statistical and historical interest? Horrors.... After all when the Russians send us replacement pages for their great encyclopedia, we add them but keep the old ones unlike (I think) the librarians in Moscow, Minsk and Leningrad."

Bill had a passionate commitment to the health of the federal statistical system and to the intelligent use of statistics in the formation and implementation of public policy. It was this commitment to the federal statistical system that was the beginning of my close collegial relationship with Bill. In 1969 he enticed me to become a member of a National Research Council panel on Problems in Census Enumeration, one of the first panels to investigate issues related to the census undercount. Although I did not know it at the time, this activity flowed into Bill's work with Allen Wallis on the President's Commission on Federal Statistics and then to the establishment of the Committee on National Statistics (CNSTAT) in 1971. Bill became its first chairman. The stamp he put on the Committee, as with all activities he was involved in shaping, was still apparent many years later when I became a member and later chairman of CNSTAT.

The most striking thing that I think most of us will always remember about Bill was his wide ranging correspondence and the system he developed of sharing his thoughts with others through the efficient mechanism of sending copies of his letters and supporting materials to a broad and diverse set of fellow intellectual travelers. How many of us have experienced, with a mix of delight and wonder, receiving a copy of a letter addressed to someone we did or did not know, with a note in the upper right hand corner in small, but legible handwriting, "File: cc:.." and then a list of names, many of whom we may not even have heard of? Often I wondered what connection there might be between me and one or two of the others who were sharing in Bill's enlargement of our horizons.

Sometimes, of course, they were letters addressed directly to me with his characteristically precise critique of an article I had written, or a penetrating question that showed all too plainly that I had missed an important body of literature, with a parenthetical reference to a book or article that I should certainly read or, if deserved, a bit of praise for making some point that he particularly approved of or that was nicely phrased. If the letter contained some criticisms that were not only just, but suddenly made one realize one's own shortcoming, then the delicate sentence that invariably ended such a letter, "I hope you will not mind if I send a few copies of this exchange to possibly interested colleagues in addition to those listed," was read with trepidation.

*Norman Bradburn is a member of the National Opinion Research Center, University of Chicago, Chicago, Illinois 60637, USA e-mail: bradburn-norman@norc.uchicago.edu.*

Not only did Bill correspond with a wide range of people—in a cursory review of my files, I found letters to colleagues in Vienna, Novosibirsk, Canberra and Allenbach, Germany—but they covered what seems like an impossibly large set of topics—statistical, historical, bibliographical puzzles, the accuracy of income statistics, the role of advisory committees, an early use of the term "public opinion" by Trollope, the misuse of significance tests—to mention just a few that are in my files.

Although Bill's critical eye was sharp and he never was one to pull punches, he did manage to convey his comments in a delightful language that (mostly) took the sting out of his critique. For example, "Would it have been totally impractical to do the randomization before bringing those sweet women into sight of the Promised Land?" or by using a memorable metaphor, "Yet there is one bone that sticks in my throat, and I write to pick it, thus probably mixing two osteo-metaphors."

Sometimes his irritation at some persistent misuse of statistics would boil over and he would be more direct, as with the author of an article that used $p$-values to assess the importance of differences, a topic that was especially dear to his heart. "So I'm sorry that this ubiquitous practice received the accolade of use by you and your distinguished coauthors. I am thinking these days about the many senses in which relative importance gets considered. Of these senses, some seem reasonable and others not so. Statistical significance is low on my ordering." Then, ever the gentleman, he adds, "Do forgive my bluntness."

Bill also had a precise sense of language that makes the author of *Eats, Shoots and Leaves* look like an amateur. In response to my request for his opinion on someone we were thinking of hiring, he replied, "[A]s of now I do not have a crisp opinion." When I asked him to be the acting dean of the then new Harris School of Public Policy Studies, he replied, after taking a respectable amount of time to think about it, that he would do it only on the condition that he be designated the "dean pro tempore." I have always thought that this reflected a sense of senatorial dignity and courtliness that must have been part of his self-image.

Bill was modest and devoid of vanity himself. He was Apollonian rather than Dionysian. He had a wry sense of humor, but I only once remember him really laughing. One day during a conversation about an academic appointment while he was dean of Social Sciences, I told him the Max Beerbohm story about Enoch Soames, a minor English poet who sold his soul to the devil in exchange for immortality. To prove that he had kept his side of the bargain, the devil sends Soames forward a hundred years in time to look himself up in the catalog of the British Library. There he finds the entry: "Enoch Soames, an imaginary character in a short story of Max Beerbohm." Bill laughed uproariously. I think the story must have epitomized for him all the vanities and uncertainties among the faculty with which academic administrators have to deal.

After his retirement, Bill remained active professionally. He especially enjoyed being a visiting fellow at the Program Evaluation and Methodology Division of the General Accounting Office, which gave him a new venue in which to pursue his life long interest in improving the uses of statistics by the federal government. His legacy in national statistics, as well as in the discipline and in the University, is manifest in all of us, his colleagues, his students and his myriad friends. We shall miss the conversations and the correspondence, but will ever remember him as an extraordinary human being.